# Controlled rotation of electrically injected spins in a non-ballistic spin field-effect transistor

F. Eberle, D. Schuh, B. Grünewald, D. Bougeard, D. Weiss, and M. Ciorga[*]

*Institute for Experimental and Applied Physics,*
*University of Regensburg, D-93040 Regensburg, Germany*

## Abstract

Electrically controlled rotation of spins in a semiconducting channel is a prerequisite for the successful realization of many spintronic devices, like, e.g., the spin field effect transistor (sFET). To date, there have been only a few reports on electrically controlled spin precession in sFET-like devices. These devices operated in the ballistic regime, as postulated in the original sFET proposal, and hence need high SOC channel materials in practice. Here, we demonstrate gate-controlled precession of spins in a non-ballistic sFET using an array of narrow diffusive wires as a channel between a spin source and a spin drain. Our study shows that spins traveling in a semiconducting channel can be coherently rotated on a distance far exceeding the electrons' mean free path, and spin-transistor functionality can be thus achieved in non-ballistic channels with relatively low SOC, relaxing two major constraints of the original sFET proposal.

Datta and Das have suggested a spintronic device,[1] named subsequently the spin-field-effect-transistor (sFET),[2] exploiting the Rashba effect[3] to electrically control the rotation of spins in a ballistic channel. The requirement of ballistic transport means that the length of the sFET's channel has to be shorter than the electrons' mean free path, typically around 1 µm. To be able to rotate spins by π on such a small distance, systems with large spin-orbit coupling must be used, which limits the selection of suitable materials. Being able to use a diffusive channel in a spin transistor would therefore lift the constraint of large SOC. What is more, whereas the electrical injection of spins into ballistic channels proved to be very challenging,[4] electrical spin injection into diffusive systems has been understood quite well[5,6] and realized in various materials.[7–11] However, although a gate-controlled Rashba effect has been demonstrated in both ballistic and diffusive two-dimensional electron gas (2DEG) channels,[13–16] gate-controlled precession of electrically generated spins has been demonstrated so far in ballistic 2DEG channels only,[17–20] and only one of these experiments came close to the originally proposed geometry, lacking however one-dimensional (1D) confinement.[17]

A diffusive system, which we employ in our devices, seems to be detrimental at first sight. In a diffusive channel, electrons undergo a series of scattering events that change the direction and the strength of the spin-orbit effective magnetic field $\boldsymbol{B}_{so}$. This randomizes the spin orientations due to the Dyakonov-Perel mechanism of spin relaxation,[21] reducing the spin-related signals. Even worse, the *k*-dependence of $\boldsymbol{B}_{so}$ prohibits a coherent rotation of a spin ensemble. Various approaches have been suggested to realize the sFET in diffusive systems.[22,23] One of them invokes the persistent spin helix (PSH) symmetry, with a unidirectional orientation of $\boldsymbol{B}_{so}$, independent of the wave vector.[23,24] To achieve the PSH symmetry one has to carefully tailor material parameters to match the strength of RSOC with Dresselhaus spin-orbit coupling (DSOC).[24–26] Alternatively, it was demonstrated that narrowing the width of a diffusive channel leads to similar effects as PSH symmetry.[27] Reducing the possible number of transversal *k*-vectors leads to a unidirectional $B_{so}$[15], which results, e.g., in an increased value of spin relaxation times, [28–31], and also could lead to coherent spin precession.[32,33]

In this paper, we show that the unidirectional $B_{so}$ in narrow channels enables coherent rotation of electrically injected spins, which can be controlled by an external electric gate. Our experiments were performed on an array of narrow wires defined in an (In,Ga)As quantum well. We electrically inject spins into these wires using a (Ga,Mn)/GaAs spin Esaki diode.[10,11,34] With an external, in-plane oriented magnetic field $B_{in}$, we orient the spins along the desired direction and observe spin precession depending on the total effective magnetic field, i.e. the superposition of $B_{so}$ and $B_{in}$. Utilizing a gate placed on top of the array, we tune the spin signal and show that we can rotate the orientation of the spin ensemble by $\pi$ on distances much larger than the mean free path of the system.

In Fig. 1 we show a schematic of our nonlocal experimental devices, all fabricated from a single heterostructure (see Supporting Information), and an SEM picture of a typical sFET device (still without a gate). The transport channel consists of an array of thirteen ~1 µm wide wires etched out of an $In_{0.09}Ga_{0.91}As$ quantum well and placed between the spin injecting and detecting contacts, constituting spin source and spin drain, respectively. The width of the wire is slightly larger than the typical electron mean free path of ~ 0.7 µm (see Supporting Information) and much smaller than the measured spin diffusion length $\lambda_s$. Recently, we have demonstrated an enhanced $\lambda_s$ in channels oriented along [100] direction,[31] indicating that $B_{SO}$ became unidirectional. Here, we investigate devices with channels oriented along the [110] crystallographic direction, with and without a top gate. The part of the spin accumulation generated in the 2DEG underneath the spin *injector* diffuses towards the *detector* contact, where it is detected as a nonlocal voltage $V_{nl}$. The orientation of the injected spins is set by the magnetization direction of the ferromagnetic (Ga,Mn)As contact, which for low $\boldsymbol{B}_{in}$, is oriented typically along an easy magnetic axis, i.e., the [100] or [010] direction,[35] and at larger $\boldsymbol{B}_{in}$, along $\boldsymbol{B}_{in}$. The injected spins move in the presence of the total magnetic field being the sum of $\boldsymbol{B}_{in}$ and the spin-orbit effective magnetic field $\boldsymbol{B}_{SO}$. A gate placed above the array of wires controls $\boldsymbol{B}_{SO}$, as discussed later.

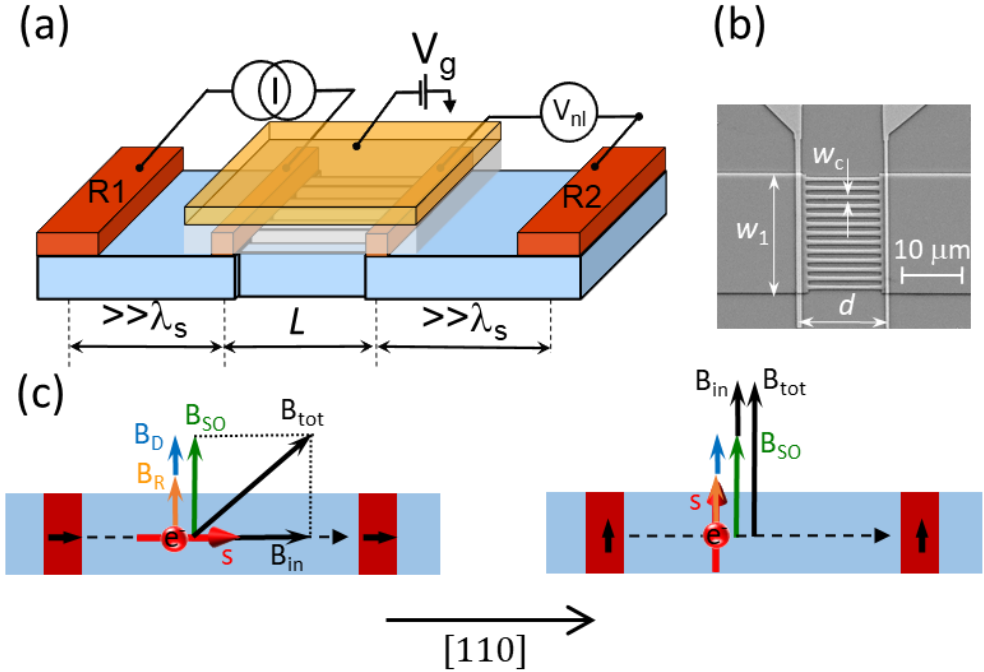

**Figure. 1.** (a) Schematic of the experimental device. Electrical spin injection takes place as a result of driving a charge current between the narrow injector contact placed close to the middle of the mesa and the reference contact R1 on the left side of the mesa. The nonlocal voltage $V_{nl}$ is measured between the detector and the reference contact R2. Gate voltage $V_g$ controls the spin-orbit coupling in gated samples. Typical sizes of injector and detector contacts are 500 nm and 700 nm, respectively; the investigated samples had channel lengths L=7, 11 and 15 µm. (b) SEM picture of one of the devices without a gate. The array of narrow wires is placed between the narrow spin contacts. (c) Cartoons showing the relative alignment of spin-orbit field $\boldsymbol{B}_{SO} = \boldsymbol{B}_R + \boldsymbol{B}_D$, an external in-plane magnetic field $\boldsymbol{B}_{in}$ and the injected spin $\boldsymbol{s}$ in narrow channels along the [110] direction for spins parallel (left) and perpendicular (right) to the wire direction.

To quantify the spin rotation we need a closer look at the SOC. In heterostructures made of zinc-blende semiconductors, the effective magnetic field reads $\boldsymbol{B}_{SO} = \boldsymbol{B}_R + \boldsymbol{B}_D$, where $\boldsymbol{B}_R$ is the Rashba SOC,[3] originating from structure inversion asymmetry, and $\boldsymbol{B}_D$ is the Dresselhaus term,[25] originating from bulk inversion asymmetry (see Supporting Information).[26,36] For $\boldsymbol{k}$-vectors parallel to the [110] direction, $\boldsymbol{B}_D$ and $\boldsymbol{B}_R$ are co-linear and perpendicular to $\boldsymbol{k}$. The field strength for this direction is given by $B_{SO} = \frac{2|\boldsymbol{k}|}{g\mu_B}(\tilde{\beta} + \alpha - \beta_3)$, where $g$ is the electron's effective $g$-factor, $\mu_B$ the Bohr magneton, $\alpha$ is the Rashba parameter, describing the strength of RSOC, and $\tilde{\beta} = \beta_1 - \beta_3$, with $\beta_1, \beta_3$ being parameters describing the strength of linear and cubic DSOC, respectively. Sketches in Fig. 1 show the relative orientation of all involved $\boldsymbol{B}_{tot}$ components, assuming a unidirectional $\boldsymbol{B}_{SO}$ for the narrow transport channel. For the chosen geometry, $\boldsymbol{B}_{SO}$ is always perpendicular to the channel. When spins oriented along the channel are injected, this perpendicular component of $\boldsymbol{B}_{tot}$ drives the precession of spins around $\boldsymbol{B}_{tot}$ during diffusion. Assuming the same sign of $\alpha$ and $\tilde{\beta} - \beta_3$, Rashba and Dresselhaus's contributions add up for $\boldsymbol{k} \parallel [110]$, leading to an increased value of $\boldsymbol{B}_{SO}$.[37]

First, we characterize spin precession via the non-local voltage $V_{nl}$ measured at the detector for an ungated device with a channel length of L= 7 μm as a function of $\boldsymbol{B}_{in}$. Fig. 2 shows results for $\boldsymbol{B}_{in}$ oriented either along the contacts or along the channel. For $|\boldsymbol{B}_{in}| > 200$ mT, the magnetization in the spin-injecting contacts aligns along $\boldsymbol{B}_{in}$, and so do the injected spins. For small $|\boldsymbol{B}_{in}|$, we observe the switching of $V_{nl}$, reflecting magnetization reversal in the spin contacts. For sweeps with $\boldsymbol{B}_{in} \parallel [1\bar{1}0]$ we observe a typical spin-valve pattern[7–11] [Fig. 2(b)], resulting from switching between parallel and antiparallel alignment of magnetization in the spin contacts. As the shape anisotropy of (Ga,Mn)As contacts is not high enough to orient the magnetization along the ⟨110⟩ directions, we assume that during the switching the magnetizations in both contacts, and the injected spins, are oriented along the [100] direction, the easy cubic magnetic axis.[35] The amplitude of the spin-valve signal $\Delta V_{nl}^{[100]}$ is a measure of the generated spin accumulation (see Supporting Information).

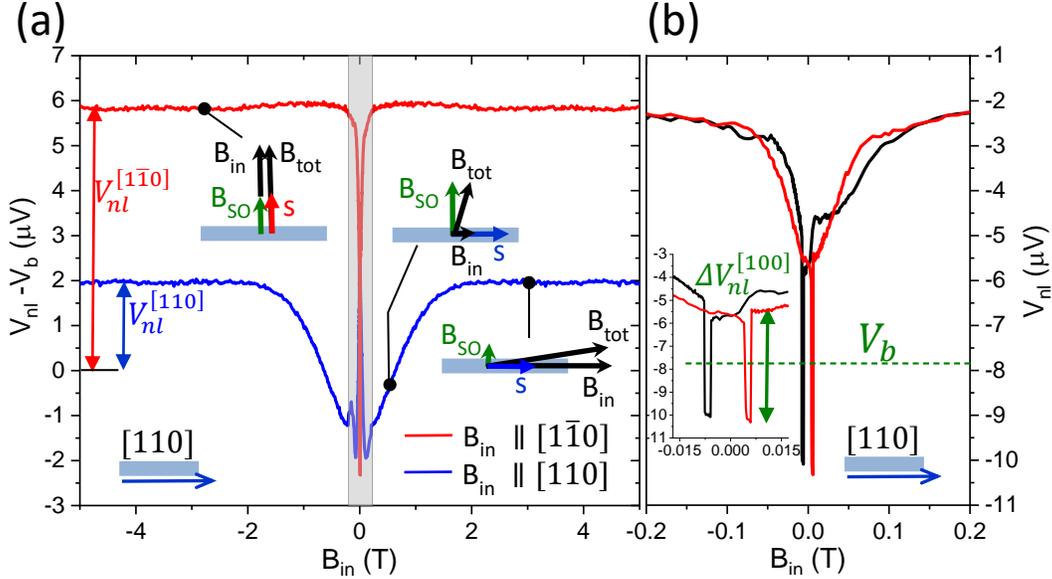

**Figure 2.** (a) Nonlocal voltage for *L*=7 μm measured as a function of $\boldsymbol{B}_{\text{in}}$, for $\boldsymbol{B}_{\text{in}}$ along the channel (blue) and along the spin contacts (red). For simplicity, only an up sweep is shown. The background voltage $V_b(B)$ has been removed from the data (see Supporting Information), with $V_b$ at $B=0$ estimated from the low-field data [see (b)]. Insets show schematically the alignment of spin *s*, the spin-orbit field $\boldsymbol{B}_{\text{SO}}$ and the total field $\boldsymbol{B}_{\text{tot}} = \boldsymbol{B}_{\text{SO}} + \boldsymbol{B}_{\text{in}}$ for high and low $\boldsymbol{B}_{\text{in}}$. (b) Raw data of the nonlocal signal for $\boldsymbol{B}_{\text{in}} \parallel [1\bar{1}0]$, within the *B*-field regime marked in grey in (a). Both up (red) and down (black) sweeps are shown. A typical spin valve pattern is observed as a result of switching between antiparallel and parallel orientation of magnetization in spin contacts. The downward bending of the signal upon decreasing magnetic field reflects rotation of magnetization towards the easy [100] direction, the starting point of the switching process. The latter is shown in the inset. The background voltage $V_b$ at $B=0$ [subtracted from the raw data in (a)] is determined as the voltage around which the switching takes place (see Supporting Information).

Let us now focus on the B-dependence of $V_{nl}$ for $|\boldsymbol{B}_{\text{in}}| > 200$ mT, shown in Fig. 2(a). For $\boldsymbol{B}_{\text{in}}$ oriented along the contacts (perpendicular to the channel direction), the spin signal $V_{nl}^{[1\bar{1}0]}$ does not depend on $\boldsymbol{B}_{\text{in}}$ in this field range [red trace in Fig. 2(a)]. Spins are there oriented along $\boldsymbol{B}_{\text{tot}}$, collinearly with $\boldsymbol{B}_{\text{SO}}$. For $\boldsymbol{B}_{\text{in}}$ aligned along the [110] channel (blue trace), however, we measure a negative $V_{nl}$ at low $\boldsymbol{B}_{\text{in}}$, which changes sign with increasing $\boldsymbol{B}_{\text{in}}$ and saturates for $|\boldsymbol{B}_{\text{in}}| \gtrsim 2$ T. We can understand the negative signal if we assume that the spins precess around $\boldsymbol{B}_{\text{tot}} \cong \boldsymbol{B}_{SO}$ for $\boldsymbol{B}_{\text{in}} \cong 0$. As a result, the spins arrive at the detector with an antiparallel orientation to its magnetization. With increasing $\boldsymbol{B}_{\text{in}}$, $\boldsymbol{B}_{\text{tot}}$ approaches $\boldsymbol{B}_{\text{in}}$ and the spin signal saturates at $V_{nl}^{[110]}$. Further, we point out a striking difference in the amplitude of the signal for spins aligned along the contacts and along the channel, i.e., along $[1\bar{1}0]$ and $[110]$ directions, respectively, where $V_{nl}^{[1\bar{1}0]} > V_{nl}^{[110]}$ holds. Although different injection efficiencies $P$ for the two directions might cause differences, this effect should be rather small.[38] Instead, we attribute this difference to the anisotropic spin relaxation time, connected to the anisotropic $\boldsymbol{B}_{SO}$,[37,39,40] which leads to different values of $\lambda_s$ for different spin directions.

After we have identified the spin-precession-related features in the signal, we performed gate experiments intending to control $\boldsymbol{B_{SO}}$ using an external electric field. In Fig. 3 we summarize the effect of the gate voltage on the spin transport properties. Upon increasing the gate voltage, the spin signal decreases, both for spins aligned along the contacts, i.e., along the $[1\bar{1}0]$ direction, as well as for spins along the $[100]$ direction, an easy magnetic axis of the contact. Such a behavior can be partially attributed to a decreasing sheet resistance because of the increased carrier density and mobility (see Supporting Information). As shown in Fig. 3(b), however, $\lambda_s$, extracted from the exponential decay of the spin signal, also decreases with increasing gate voltage for both spin orientations, suggesting an increasing spin relaxation rate. In the whole gate voltage range, $\lambda_{[1\bar{1}0]} > \lambda_{[100]}$ holds, confirming the observation for the ungated sample. Within the Dyakonov-Perel mechanism, $\lambda_s$ is inversely proportional to the spin-orbit coupling strength,[21,27,31] so decreasing $\lambda_s$ means increasing SOC. This observation supports our claim that a gate voltage affects SOC in our samples. The extracted spin injection efficiency $P$ does not depend on spin direction [see the upper panel of Fig. 3(b)], consistent with the assumption made earlier.

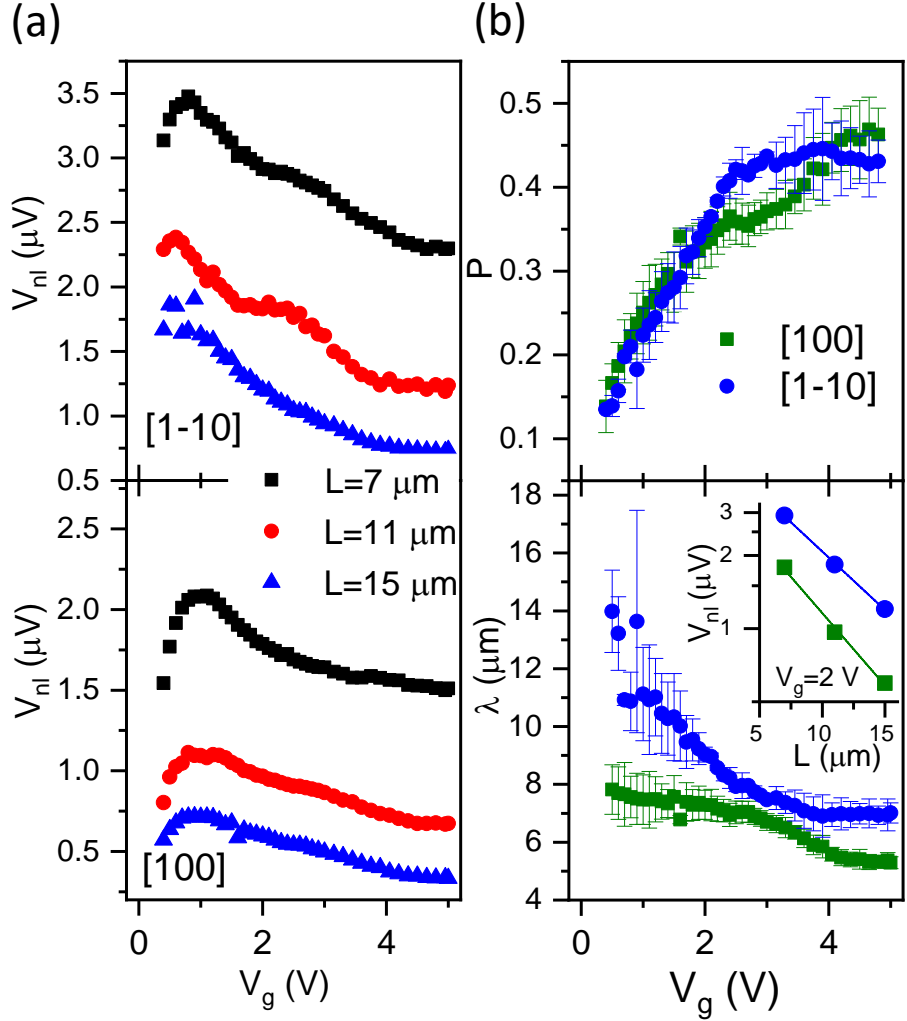

**Figure 3.** (a) Gate dependence of the spin signal $V^{[1\bar{1}0]}$ (top) and $V^{[100]} = \Delta V_{nl}^{[100]}/2$ (bottom) for spins oriented along $[1\bar{1}0]$ and $[100]$ directions, respectively. (b) Spin diffusion length $\lambda_s$ (bottom) and spin injection efficiency $P$ (top) obtained from the distance dependence shown in (a). Inset: distance dependence of signals for the selected voltage $V_g = 2$ V, showing an exponential decay, characteristic for a spin-diffusion process.

Now we discuss how the gate voltage influences the spin-precession-related features in the $B$-field sweeps. In Fig. 4(a) we show field sweeps for different gate voltages, $L$=11 μm and $\boldsymbol{s} \parallel \boldsymbol{B}_{\text{in}} \parallel [110]$, i.e., for spins oriented along the channel. The signal shape strongly depends on the gate voltage, particularly for $|\boldsymbol{B}_{\text{in}}| \lesssim 2T$. For $V_g = 1.1\,V$ the low-field signal is positive and larger than observed at high fields. Upon increasing the gate voltage, the signal first decreases and then changes sign. For $V_g \gtrsim 2.2\,V$ it resembles the shape observed for the ungated sample, shown in Fig. 2(a). The sign of the signal changes, corresponding to a change of the spins' orientation relative to the magnetization of the detector contact. These data show that the measured signal is exclusively due to the spin and its interaction with the effective spin-orbit field.

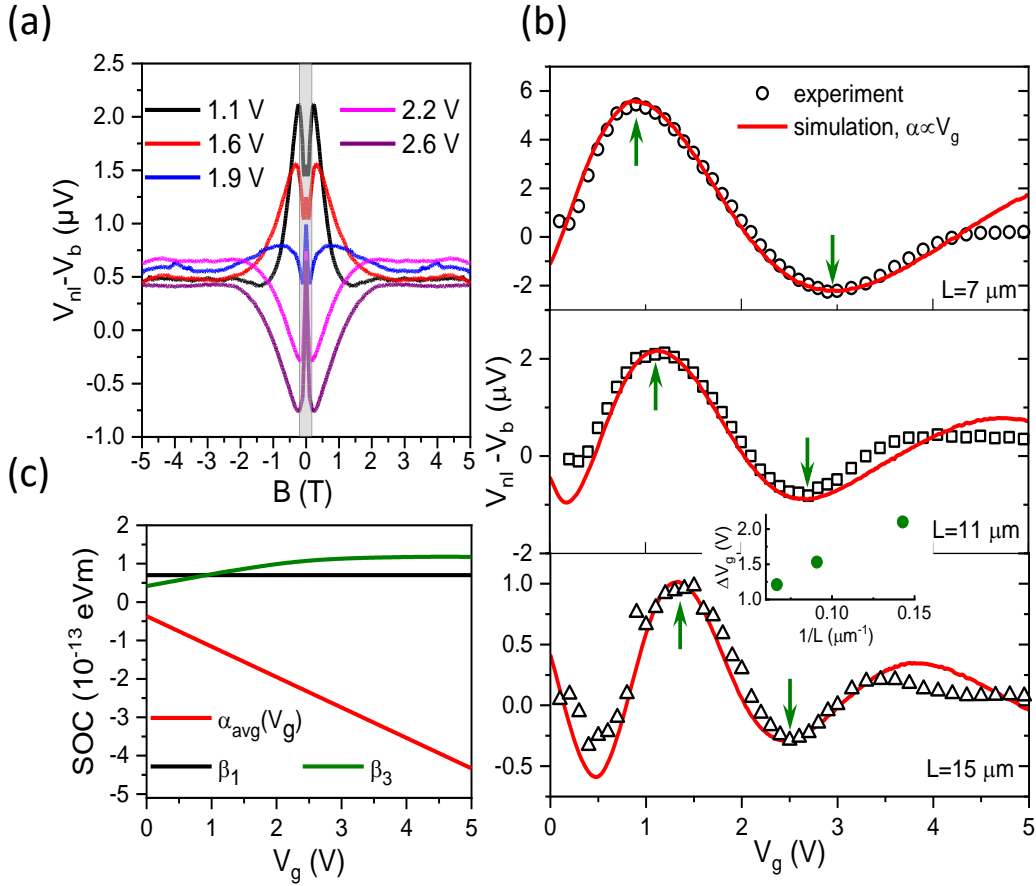

**Figure 4.** (a) Nonlocal voltage for $L$=11 μm measured as a function of $\boldsymbol{B}_{\text{in}}$ along the [110] channel for the selected gate voltages. The SOC-related feature at low fields changes as a function of a gate voltage. (b) oscillations of the spin signal for different channel lengths as a function of the gate voltage at fixed $B_{\text{in}} = 0.2$ T. Experimental data are shown as symbols. The red curves are calculated assuming a linear $\alpha(V_g)$ dependence. (c) SOC parameters used to obtain the simulated curves plotted in (b). Please note that $\beta_1 - 2\beta_3 < 0$ holds in the whole measured voltage range.

Finally, we focus on the key sFET's functionality, i.e., oscillations of the spin signal at the drain, controlled by the gate voltage. In Fig. 4(b) we show the dependence of the signal on the gate voltage, probed at fixed $\boldsymbol{B}_{\text{in}} = 0.2$ T (i.e., where magnetization switching is completed) for three different distances between the injector and detector. The signal shows oscillations in all three cases, with arrows indicating maxima and minima. Maxima (minima) correspond to spins at the detector being parallel (antiparallel) to the magnetization of the detector, resulting in a positive (negative) $V_{nl}$. The difference in gate voltage $\Delta V_g$ between maximum and minimum, equivalent to one-half of an oscillation period, corresponds to the spins being rotated by $\pi$ while traveling between injector and detector. This value decreases from $\Delta V_g \cong 2.1\ V$ for $L$=7 μm to $\Delta V_g \cong 1.2\ V$ for $L$=15 μm, i.e., the period of the oscillations decreases with an increasing channel length. The observed oscillation of the spin signal proves therefore that we can rotate the spins via a gate voltage also in a diffusive channel, i.e., we achieve spin transistor functionality.

Below we describe in more detail the observed oscillations in connection with the theoretical predictions. In a ballistic channel, following the arguments of Datta and Das, oscillations of the nonlocal signal can be described by a single cosine function $\cos[2m^*\alpha(V_g)L/\hbar^2]$, where $\alpha(V_g)$ is the voltage-dependent RSOC parameter. A corresponding full period of the oscillations is therefore connected to a change of $\alpha$ by $\Delta\alpha = \frac{\pi\hbar^2}{m^*L}$. If $\alpha(V_g)$ depends linearly on $V_g$ then $\Delta V_g \propto 1/L$. While the $1/L$ periodicity is observed in our experiment [see the inset in Fig. 4(c)], the simple cosine does not correctly describe the voltage dependence of the oscillation amplitude. Some theoretical work suggested, that for 2D channels, particular attention has to be paid to the phase of the oscillations and the amplitude.[41–43] In particular, Zainuddin et al.[41] provided a more general expression for $V_{nl}$ (see Supporting Information) taking into account higher *k*-mode contributions to the oscillations, which for spin precession angles $\theta_L > 2\pi$ is well approximated by

$$V_{nl} = V_{nl}^0 + \frac{A}{\sqrt{2\pi(\theta_L + \theta_0)}} \cos\left(\theta_L + \theta_0 + \frac{\pi}{4}\right). \quad (1)$$

Here, $\theta_L = 2m^*\alpha(V_g)L/\hbar^2$ is the angle by which spins precess on a distance $L$ and $\theta_0$ indicates any additional arbitrary phase related, e.g., to the final dimension of the ferromagnetic contacts. Although for our experimental conditions, the general formula [Eq. (S7) in the Supporting Information] must be used to reproduce the experimental data, Eq. 1 is useful to qualitatively describe some of the features observed in our experiment. According to Eq. 1, the amplitude of the oscillations is given by $\frac{A}{\sqrt{2\pi(\theta_L+\theta_0)}}$, i.e., it is expected to decrease with increasing $\theta_L$, differently than in the Datta-Das theory. As in our experiments the amplitude decreases with $V_g$ [see. Fig. 4(b)], this suggests that increasing $V_g$ increases the absolute value of $\alpha(V_g)$, which is consistent with a decreasing $\lambda_s$. The parameter $A$ can be expressed as $A = \frac{3\pi}{2} V_{nl}^{[1\bar{1}0]}$, where $V_{nl}^{[1\bar{1}0]}$ is the signal measured for spins oriented parallel to the contacts, i.e., along the $[1\bar{1}0]$ direction (see Fig. 2). The non-oscillatory part $V_{nl}^0$ can also be expressed in terms of $V_{nl}^{[1\bar{1}0]}$ as $V_{nl}^0 = \frac{V_{nl}^{[1\bar{1}0]}}{2}$, indicating that oscillations are not symmetric for the zero-spin signal. This is also observed in our data.

In Fig. 4(b) we plot theoretical curves obtained using the general expression from Zainuddin *et al.*[41] to simulate the experimental data. The crucial adjustable parameter is the voltage-dependent SOC parameter. In the simplest case, with only RSOC present, the tunability of SOC is given by the gate voltage dependence of $\alpha(V_g)$. In our samples, however, the contribution of DSOC must be also considered. Assuming a 1D-like character of electrical transport in our narrow channels, for $\mathbf{k} \parallel [110]$, $\alpha$ in the above expressions needs to be replaced by $\alpha^* = \alpha + \beta_1 - 2\beta_3$. Whereas $\beta_1$ is typically independent of the gate voltage, both $\alpha$ and $\beta_3$ are voltage dependent. $\beta_3$ depends on the carrier density $n_s$, as $\beta_3 = -\gamma \frac{\pi n_s}{2}$,[40,44] where $\gamma$ is a material-dependent bulk DSOC parameter. Generally, taking $\alpha(V_g) = \alpha(V_g = 0) + \frac{\partial \alpha}{\partial V_g} V_g$, we describe the voltage dependence of the total SOC strength along [1$\bar{1}$0] as $\alpha^*(V_g) \cong \alpha_0^* + \frac{\partial \alpha}{\partial V_g} V_g - \gamma \pi n_s(V_g)$, where $\alpha_0^* = \beta_1 + \alpha(V_g = 0) - 2\beta_3(V_g = 0)$. The linear DSOC parameter $\beta_1$ is given by $\beta_1 = -\gamma \langle k_z^2 \rangle$, where $\langle k_z^2 \rangle$ is the expected value of the squared wave number along the growth direction, describing the electron confinement in the $z$-direction, which cannot be accessed experimentally. For an infinite quantum well with a width of = 20 nm, $\langle k_z^2 \rangle = \left(\frac{\pi}{a}\right)^2 \cong 2.42 \cdot 10^{16}$ m$^{-2}$, which can be treated as an upper estimate of the $\langle k_z^2 \rangle$ for our low-band-offset structure. For a similar structure Studer *et al.*[45] calculated $\langle k_z^2 \rangle \cong 0.95 \cdot 10^{16}$ m$^{-2}$, which gives $\beta_1 \cong 0.7 \cdot 10^{-13}$ eV·m, taking $\gamma = -7.5 \cdot 10^{-30}$ eV·m$^3$. We use this value in our simulations. The dependence $n_s(V_g)$ we take from magnetotransport experiments performed on another sample, without spin contacts, fabricated from the same heterostructure (see Supporting Information). As we do not have experimental access to $\alpha(V_g)$, we use $\alpha(V_g = 0)$ and $\frac{\partial \alpha}{\partial V_g}$ as primary adjustable parameters. Other adjustable parameters are the non-oscillatory offset voltage $V_{nl}^0$, an arbitrary phase $\theta_0$ and the amplitude $A_0$, defined via $A = A_0 V_{nl}^{[1\bar{1}0]}(V_g)$, where $V_{nl}^{[1\bar{1}0]}(V_g)$ is the experimentally determined spin signal [see Fig. 2(a)].

Fig. 4(b) shows the simulated curve for each detector while Fig. 4(c) depicts the used gate dependence of the SOC parameters. The curves are obtained assuming $\alpha(V_g) \propto V_g$ and $\beta_3(V_g) = -\gamma \frac{\pi n_s(V_g)}{2}$. In general, we obtain remarkably good agreement with the experiment when taking $\alpha(V_g) < 0$ and $\frac{d\alpha}{dV_g} < 0$, with $\frac{d\alpha}{dV_g}$ ranging from $-0.9 \cdot 10^{-13}$ eVm/V for $L$= 7 μm to $-0.7 \cdot 10^{-13}$ eV·m/V for $L$=15 μm, giving an average value of $-0.8 \cdot 10^{-13}$ eV·m/V. With $\beta_1 -$ $2\beta_3 < 0$, that means that $\alpha^* = \beta_1 - 2\beta_3 + \alpha$ is also negative and that $|\alpha^*|$ increases with $V_g$. The increase of the total SOC strength with $V_g$ is consistent with the decreasing oscillation amplitude and with decreasing $\lambda_s$, measured for spins along $[1\bar{1}0]$ and $[100]$ directions [see Fig. 3(b)]. A negative $\alpha(V_g)$ is also consistent with our toy model discussed in Fig. 1, where we assumed large $\boldsymbol{B}_{SO}$ for $\boldsymbol{k} \parallel [110]$ due to Rashba and Dresselhaus spin-orbit field having the same orientation. One should note here that $\alpha(V_g = 0)$ does not have a strong effect on the oscillating part of the signal, therefore the obtained absolute values of $\alpha^*$ and $\alpha$ are inexact and must be viewed with caution. Nevertheless, the absolute value of $\alpha^*$ influences the amplitude of oscillations via $[2\pi(\theta_L + \theta_0)]^{-1/2}$ what allowed us to estimate $\alpha$ being in the order of low $10^{-13}$ eVm. Lastly, we would like to note, that the discrepancy between the simulated curves and the experimental data for $V_g > 3.5$ V in Fig. 4(b) can be explained by the following: In our simulations we assumed for simplicity that $\frac{d\alpha}{dV_g}$ is constant up to $V_g = 5\,V$, whereas it is clear from the experimental data that there are no oscillations beyond $V_g \gtrsim 4.2\,V$, suggesting that there is no modulation of Rashba parameter anymore at these voltages.

In summary, our experiments demonstrate that in narrow diffusive channels spins can be coherently rotated in a controlled manner over a channel length much longer than the elastic mean free path. This shows that functional sFETs are not limited to short ballistic channels with large SOC, as postulated in the original proposal, which significantly increases their applicability. It could be of particular importance for the sFET realization in a novel van der Waals material platform, where one typically observes short mean free paths. As the mean free path decreases with temperature, our observation could also open a way to sFET operation at elevated temperatures. Although the generalized ballistic model describes spin transport in these long channels surprisingly well, the more realistic model should consider possible spin lifetime anisotropies in the channel.

**Supporting Information**

Supporting Information: description of experimental methods, including sample fabrication and measurements; gate dependence of electric properties of the samples; discussion of spin-orbit effective magnetic field; description of background removal procedure; description of the spin-valve signal in the array geometry; description of the theoretical model of voltage-controlled oscillations.

**Corresponding Author**

*Correspondence should be addressed to mariusz.ciorga@ur.de

**Author Contributions**

The manuscript was written through contributions of all authors. All authors have given approval to the final version of the manuscript.

**Acknowledgments**

We would like to thank Prof. J. Fabian for fruitful discussions. This work was supported by the German Science Foundation (DFG) via projects No. 429749589 and ID 422 31469 5032-SFB1277 (Subproject A01).

# Supporting Information for

## Controlled rotation of electrically injected spins in a non-ballistic spin field-effect transistor

F. Eberle, D. Schuh, B. Grünewald, D. Bougeard, D. Weiss and M. Ciorga[*]

*Institute for Experimental and Applied Physics,*

*University of Regensburg, D-93040 Regensburg, Germany*

**Supporting notes**

1. Experimental methods
2. Spin-orbit coupling effective magnetic field
3. Gate dependence of the electronic properties
4. Background removal procedure
5. Spin-valve signal in the array geometry
6. Theoretical description of voltage-controlled signal oscillations

---

[*] Corresponding author: mariusz.ciorga@ur.de

## 1. Experimental methods

Sample fabrication

All investigated samples were fabricated from the same wafer, grown by molecular beam epitaxy. The wafer consists of the following layers (from top): ferromagnetic $Ga_{0.95}Mn_{0.05}As$ (50 nm), $Al_{0.33}Ga_{0.67}As$ (2 nm), $n^+$-GaAs (8 nm, $n^+= 5\times10^{18}$ cm$^{-3}$), $n^+\rightarrow n$-GaAs transition layer (15 nm,), $n$-GaAs (100 nm, $n= 6\times10^{16}$ cm$^{-3}$), $In_{0.09}Ga_{0.91}As$ layer (20 nm), $Al_{0.33}Ga_{0.67}As$ (95 nm), AlGaAs/GaAs superlattice buffer (500 nm) and (001)-oriented GaAs substrate. The top four layers form a spin Esaki diode. The two-dimensional electron gas (2DEG) is confined in the $In_{0.09}Ga_{0.91}As$ quantum well and charge carriers are provided by a Si δ-doped layer in the $Al_{0.33}Ga_{0.67}As$ region. As the (Ga,Mn)As layer is degenerately $p$-doped, it is necessary to increase the Si doping in the $n^+$-layer to ensure the Esaki diode functionality. This is achieved by the so-called pseudo-δ-doping, meaning that the growth process of the 8 nm thick n$^+$-GaAs layer using continuous Si flux is stopped every 1.6 nm for 10 s, to accumulate Si dopants. This allows n$^+$-doping higher than $1\times10^{19}$ cm$^{-3}$, making the $p$-$n$-junction more symmetric and thus lowering its resistance. Narrow transport channels are defined using electron-beam lithography (EBL) and ion beam etching. Narrow, 500 nm, and 700 nm wide, ferromagnetic electrodes are defined using EBL, followed by evaporation of 100 nm/10 nm Au/Ti contacts. Different widths of injector and detector contacts ensure different switching fields of the contacts during magnetoresistance measurements. In the region between the contacts, the top three layers were partially etched away using wet chemical etching with an acetic-acid-based solution (5 $C_2H_4O_2$:$H_2O_2$:5 $H_2O$) to limit the lateral transport to the 2DEG. For the samples described in this paper, the etching depth was between 70 and 75 nm. The electric gates were defined using EBL, followed by a plasma-enhanced chemical vapor deposition of a 10 nm thick layer of $SiO_2$ layer at 80 °C, atomic layer deposition of a 100 nm thick layer of $Al_2O_3$ at 120 °C and finally, evaporation of Ti/Au metal contacts (2 nm/20 nm). The thickness of the metal electrode is chosen such that it forms a homogeneous layer, yet is still transparent enough to allow external illumination, which is required to fully populate the 2DEG channel at cryogenic temperatures.

Measurements

Electrical magnetoresistance measurements were performed primarily in the nonlocal configuration [see Fig. 1(a) in the main text], i.e., with no charge current in the detector circuit. The measurements were performed with an ac current of 4 µA RMS applied between the injector contact and the reference contact R1 [see Fig. 1(a)]. The nonlocal voltage drop $V_{nl}$ was measured directly between a detector contact and the reference contact R2. The samples were mounted in a cryostat allowing sample rotation around the axes normal to the plane, thus enabling to change the orientation of the in-plane magnetic field. Magnetic field sweeps were performed while sweeping an external magnetic field between $B = 5\,\text{T}$ and $-5\,\text{T}$ in both field directions. For magnetotransport measurements, the sample was mounted in a holder allowing out-of-plane orientation of the magnetic field. Here an ac current of 100 nA RMS was used.

## 2. Spin-orbit coupling effective magnetic field

In heterostructures made of zinc-blende semiconductors, the effective spin-orbit magnetic field reads $\boldsymbol{B}_{\text{SO}} = \boldsymbol{B}_{\text{R}} + \boldsymbol{B}_{\text{D}}$, where $\boldsymbol{B}_{\text{R}}$ is the Rashba spin-orbit coupling (SOC),[1] originating from the structure inversion asymmetry, and $\boldsymbol{B}_{\text{D}}$ is the Dresselhaus term,[2] originating from bulk inversion asymmetry.[3,4] For heterostructures grown along the [001] direction, using the coordinate system $x \parallel [1\bar{1}0]$, $y \parallel [110]$, $z \parallel [001]$, it holds

$$\boldsymbol{B}_{\text{R}} = \frac{2\alpha}{g\mu_B}(k_y, -k_x) \tag{S1}$$

where $\alpha$ is the Rashba parameter, describing the strength of RSOC, $g$ is the electron's effective $g$-factor, and $\mu_B$ the Bohr magneton. For all possible $k$-vectors, $\boldsymbol{B}_{\text{R}}$ is always perpendicular to $\boldsymbol{k}$.

Dresselhaus's contribution can be described as $B_D = B_{D1} + B_{D3}$, where $B_{D1}, B_{D3}$ describe, respectively, the terms linear and cubic in $k$. They can be expressed as

$$B_{D1} = \frac{2\tilde{\beta}}{g\mu_B}(k_y, k_x) \tag{S2}$$

and

$$B_{D3} = \frac{2\beta_3}{g\mu_B}|k|(\sin 3\theta, -\cos 3\theta) \tag{S3}$$

where $\theta$ is the angle between the $k$-vector and the $x$-axis, and $\tilde{\beta} = \beta_1 - \beta_3$, with $\beta_1, \beta_3$ being parameters describing the strength of linear and cubic contributions, respectively. For $k$-vectors parallel to either $[1\bar{1}0]$ or $[110]$ directions, $B_D$ and $B_R$ are co-linear and perpendicular to $k$. Their alignment with respect to the channel direction and the external in-plane field $B_{in}$ is shown schematically in sketches in Fig. 1(c) of the main text. The field strength for these directions is given by $B_{SO} = \frac{2|k|}{g\mu_B}(\tilde{\beta} - \alpha - \beta_3)$ and by $B_{SO} = \frac{2|k|}{g\mu_B}(\tilde{\beta} + \alpha - \beta_3)$, respectively.

### 3. Gate dependence of carrier density and mobility.

We present here the results of magnetotransport measurements, used to characterize the given heterostructure for gate-voltage dependence of the electronic properties, namely the carrier density, the electrons' mobility, and the mean free path. The measurements were performed on a reference sample, prepared from the same heterostructure as the spin injection devices. The reference sample was shaped into a 20 µm wide Hall bar with an array of 1 µm wide stripes between the longitudinal voltage probes, similar to the spin injection devices. The entire channel is covered by a gate electrode, which is separated from the channel by 20 nm of $SiO_2$ and 100 nm of $Al_2O_3$. Measurements are performed using standard lock-in techniques with an excitation current of 100 nA. An external out-of-plane magnetic field gives rise to Shubnikov-de Haas oscillations in the longitudinal resistance $\rho_{xx}$. From the period of the Shubnikov- de Haas oscillations, the sheet carrier density $n_s$ was derived for each applied gate voltage [Fig. S1(a)]. The electrons' mobility and the mean free path [Fig. S1(b)] were then calculated as, respectively, $\mu = 1/(e n_s R_s)$ and $l_e = \hbar \mu_e \sqrt{2\pi n_s}/e$, where $R_s = \rho_{xx}(B=0)$ is the sheet resistance. The carrier density increases linearly with gate voltage up to $V_g \cong 2.5$, but saturates for voltages $V_g \gtrsim 3.5$. Whereas a linear dependence is expected from the classical field effect based on a capacitor model, the observed saturation of $n_s$ can be explained by the presence of trap states at the interfaces, which get populated instead of a quantum well. Gate dependence of $n_s$ is taken into account in the simulations curves shown in Fig. 4(b) in the main text via Dresselhaus parameter $\beta_3$.

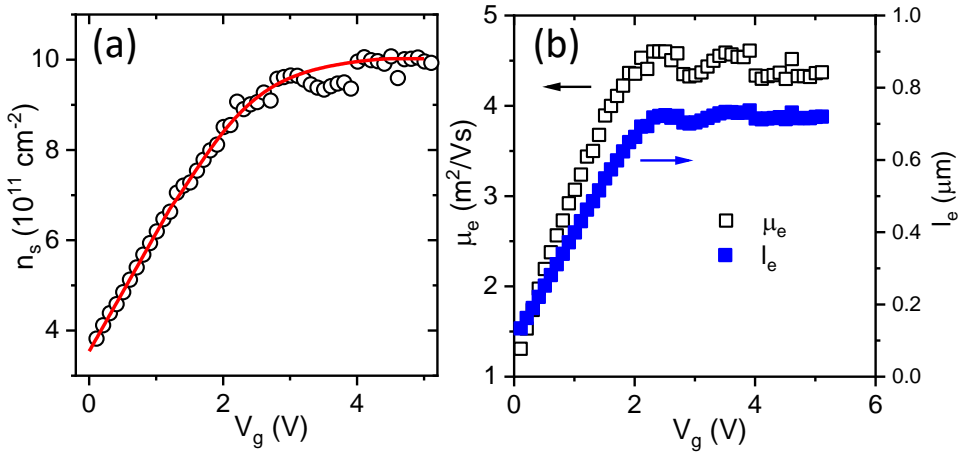

**Figure S1.** Gate-voltage dependence of (a) the carrier density $n_s$ and (b) the electrons' mobility $\mu_e$, and the electrons' mean free path $l_e$ obtained from the magnetotransport measurements on the reference sample. The red line in (a) is an interpolation curve, used to calculate the gate voltage dependence of the Dresselhaus parameter $\beta_3$, shown in Fig. 4(c) of the main text. It was used for the simulation of voltage-controlled spin precession described in the main text.

## 4. Background removal procedure

Here, we discuss how we process the raw data obtained during magnetic field sweeps to remove a field-dependent background $V_b(B)$ in the nonlocal voltage, to obtain plots shown in Fig. 2(a), (c), and Fig. 4(a) in the main text. In Fig. S2, we plot a raw nonlocal voltage $V_{nl}(B)$ measured for the channel along [110] direction for both **B** ∥ [110] and **B** ∥ [1$\bar{1}$0]. We plot both up and down sweeps of the magnetic field over the full range of $B = \pm 5$ T (a) and in a narrow range around $B = 0$ (b). We see that for both **B** directions, the up and down sweeps are identical, except in the small magnetic field range around zero [(Fig. S2(b)]. There, hysteretic behavior is observed, reflecting magnetization switching in the (Ga,Mn)As contacts. For **B** ∥ [110], the up and down sweeps fall on top of each other for $|B| \gtrsim 0.15$ T (see red arrows), indicating that the magnetization follows the external field above this field value. For **B** ∥ [1$\bar{1}$0] it happens for smaller $B$ (blue arrows), indicating that the [1$\bar{1}$0] direction is magnetically easier than [110]. At high magnetic fields, we observe a strong $V_{nl}(B)$ dependence, which we attribute mostly to a field-dependent background $V_b(B)$, not related to spin.

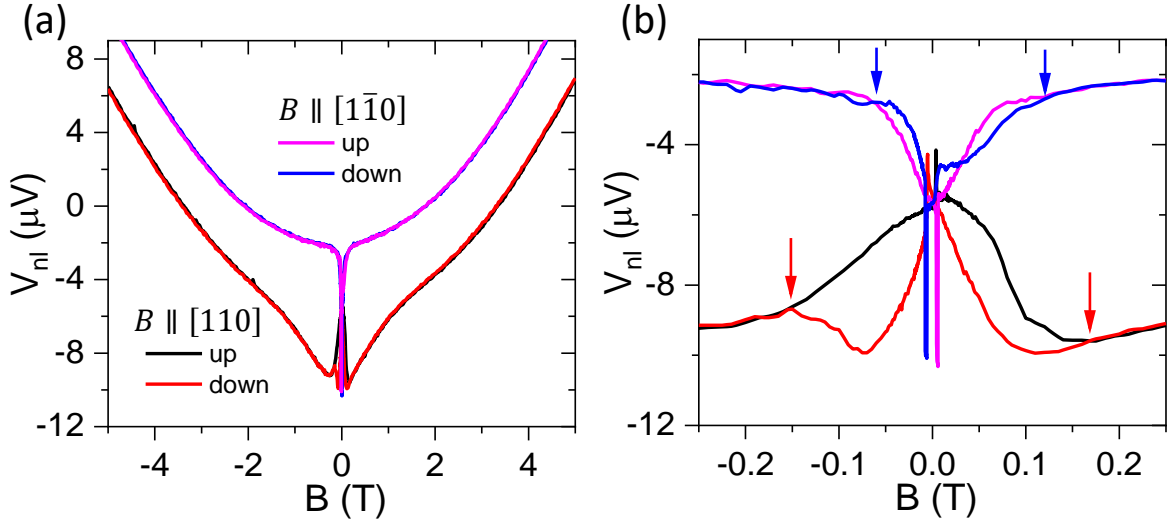

**Figure S2.** (a) Nonlocal voltage measured with the device channel along the [110] direction, while sweeping $B$ along the [110] direction, i.e., along the channel, and along [1$\bar{1}$0], i.e., along the long axis of the spin contacts. Both up and down field sweeps are shown. (b) as in (a), but showing a small range of $B$ around $B$=0. Arrows mark $B$-values, where the up and down sweeps merge with each other, indicating that from these values on, magnetizations in both spin contacts align along $B$.

In Fig. S3 we show the procedure used to remove this background to obtain the dependence of the spin signal on $B$. As an example, we show how we remove the spin-independent background for a down sweep of the magnetic field for $B \parallel$ [110], i.e., parallel to the channel. In (a) the raw data is shown. In the first step, we symmetrize the signal with respect to $B$. We base the assumption of a symmetric $V_{nl}(B)$ on our simple toy model discussed in Fig. 1 of the main text; see also the insets in the Fig. S3(b). One can clearly see that the angle between $s$ and the $B_{\text{tot}}$ is the same for positive and negative B. From this, $V_{nl}(B) = V_{nl}(-B)$ follows. We symmetrize the signal by taking $V_{nl}(B) = [V_{nl}(B) + V_{nl}(-B)]/2$. We apply this procedure only to the data for $B > 0.2$ T, in order not to affect the switching pattern. The obtained signal is plotted in (b) as the blue line. To remove the field-dependent background we use a polynomial of the fourth grade for $B > 2.5$ T, plotted as a dashed line. We subtract it from the blue curve and the resulting curve is plotted in (c). Now we shift the curve to match it with the original curve for $B = 0$. The result is plotted as the green curve in (d). Then we subtract from this curve the background voltage $V_b(B = 0) \approx -8\ \mu\text{V}$, and as a result, we obtain the final curve (blue), plotted in Fig. 2(b) of the main text.

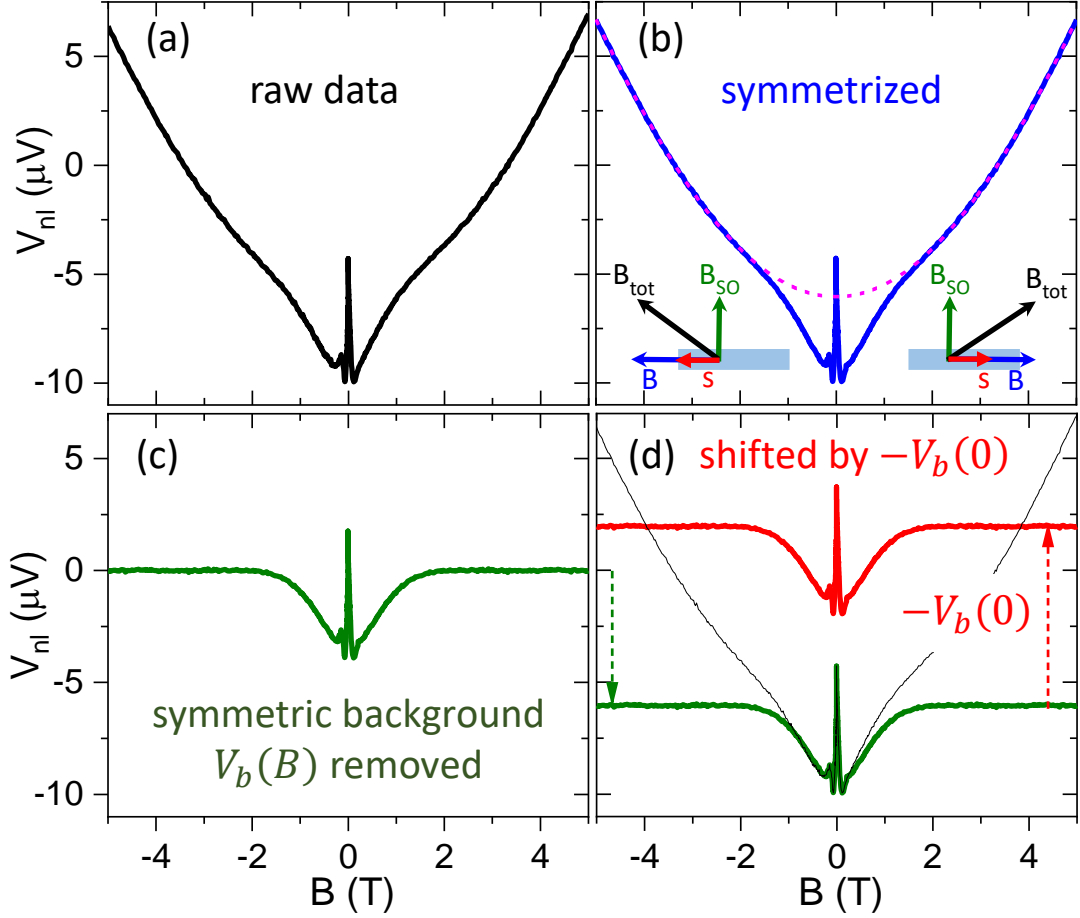

**Figure S3.** (a), Raw data taken during a down sweep of $\boldsymbol{B}$ for the [110] channel. (b) The symmetrized curve from (a) [blue solid line] and the corresponding parabolic background $V_b(B)$ [magenta dashed curve]. Insets show configurations of the total field $\boldsymbol{B}_{\text{tot}}$ (black arrow) and spin $s$ (red arrow) for a given spin-orbit field $\boldsymbol{B}_{\text{SO}}$ (green arrow) for $B > 0$ and $B < 0$. The curve obtained by subtracting the background from the symmetrized curve is plotted in (c). (d) The green curve corresponds to the curve from (c) shifted downwards (by the green dashed arrow) to match with the raw data (thin black curve) at $\boldsymbol{B} = 0$. The red curve is obtained by subtracting the background voltage at $\boldsymbol{B} = 0$, estimated as $V_b(0) \approx -8\ \mu\text{V}$, from the green curve. This is the final curve, plotted in Fig. 2(a) in the main text.

We estimate the background voltage $V_b(B = 0)$ based on the spin valve (SV) switching pattern between the voltage values corresponding to parallel ($V_P$) and antiparallel ($V_{AP}$) magnetization configurations in the injector and detector contacts. In Fig. S4 we show up and down field sweeps for three different orientations of the magnetic field. Additionally to the data for sweeps along the [110] and [1$\bar{1}$0] directions, we show also measurements for sweeping the field along the [100] direction, which is a cubic (global) magnetic easy axis for (Ga,Mn)As contacts.[5] We observe a clear SV pattern in two of those cases, i.e. for **B** ∥ [1$\bar{1}$0] and for **B** ∥ [110]. One has to note here, that for narrow (Ga,Mn)As contacts the shape anisotropy is not high enough to orient the magnetization along the ⟨110⟩ directions. We thus assume that for **B** ∥ [1$\bar{1}$0] the switching occurs when magnetization in both contacts is oriented along the easy cubic magnetic axes, i.e., along the [100] direction.[5] The spin valve signal is defined as $\Delta V_{SV} = V_P - V_{AP}$, whereas the background voltage is determined by taking $V_b(B = 0) = (V_P + V_{AP})/2$.

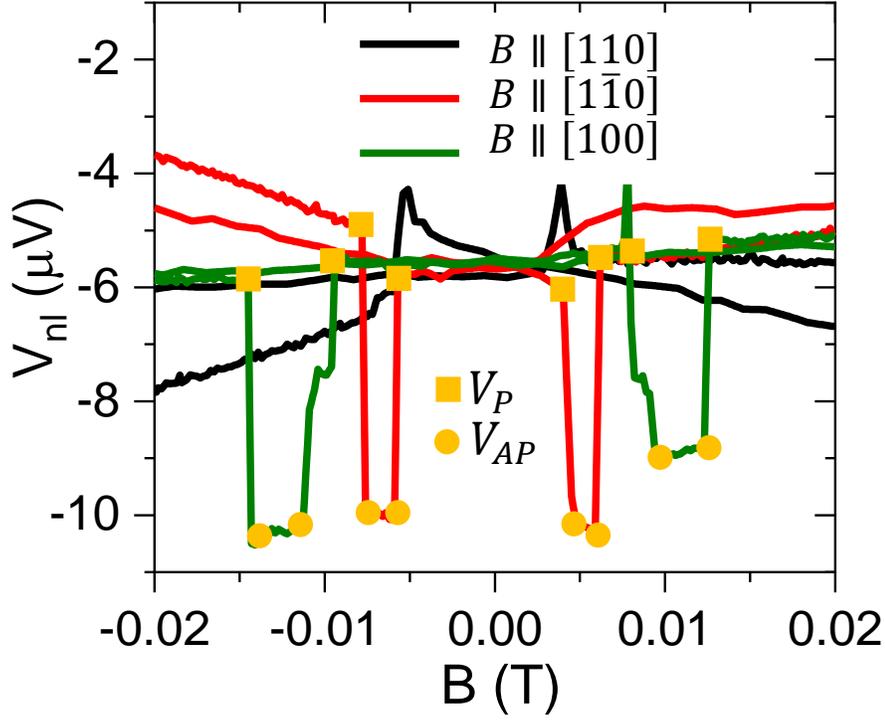

**Figure S4.** Magnetic field sweeps (raw data) around $B = 0$ for three different directions of the magnetic field. Orange squares and circles indicate field values, where switching between, respectively, parallel (P) and antiparallel (AP) magnetization configuration occurs in spin contacts. Squares (circles) mark the level of the nonlocal voltage $V_P$ ($V_{AP}$) for parallel (antiparallel) configuration. The background voltage $V_b(B = 0)$ is estimated as $V_b = (V_P + V_{AP})/2$.

The SV signal for **B** ∥ [1$\bar{1}$0] is comparable to the one for **B** ∥ [100], which is consistent with the claim that in both cases the magnetization at low **B** is oriented along the [100] direction,[5] and so is the orientation of spins injected at low **B**. We estimate then $V_b(B = 0) \approx -8\ \mu V$ by averaging over the corresponding switching values. For **B** ∥ [100] we take into account the down sweep only, as $\Delta V_{SV}$ for the up sweep clearly differs from the other ones, which may indicate that the parallel configuration was not obtained for that sweep direction. We neglect also the **B**-dependence of $V_b$ in the shown range of the magnetic field and assume that $V_b(B=0)$ is the same for both field directions. We would like to note, that this approximation is only valid at small **B**, as, in general, $V_b(B)$ for larger fields can depend on the direction of **B**.

A similar procedure is also used to remove the background for the data obtained in the gated measurements, plotted in Fig. 4(a) in the main text.

## 5. Spin-valve signal in the array geometry

For the employed array geometry, the spin valve signal $\Delta V_{SV}$, which is a measure of spin accumulation generated in the channel, is approximated by[6]

$$\Delta V_{SV} \cong \frac{4P^2 I R_{s,2} \lambda_2}{w_2} \frac{1}{\left(1 + \frac{R_{s,2}\lambda_2 w_1}{R_{s,1}\lambda_1 w_2}\right)^2} \exp\left(-\frac{L}{\lambda_2}\right) \qquad (S4)$$

Here, $P$ is the spin injection efficiency, $R_s$ the sheet resistance, $\lambda$ the spin diffusion length, $w$ the total width of the transport channel, and $I$ is the injection current. The subscript "2" ("1") indicates the region between (outside) the injector and detector contacts. For an array of $N$ narrow wires of a width $w_c$, $w_2 = N w_c$. From the distance dependence of $\Delta V_{SV}$, spin diffusion length in the wires $\lambda_2$ can be extracted ($\lambda_s$ in the main text). $\lambda_1$ can be extracted from SV measurements on the standard, i.e., non-array, geometry with $w_2 = w_1$. After determining sheet resistances $R_{s1,2}$ from the magnetotransport measurements, spin injection efficiency $P$ can be then also extracted from eq. S4. Gate-voltage dependence of $P$ and $\lambda_s$ for spins oriented along $[1\bar{1}0]$ and $[110]$ is shown in Fig. 3 in the main text.

## 6. Theoretical description of voltage-controlled signal oscillations

For simulations of the observed signal oscillations we used the theoretical model of Zainuddin *et al.*[7]. They provided a general formula taking into account higher $k$-mode contributions to the oscillations

$$V_{nl}^s = \int_{-k_F}^{k_F} \frac{dk_y}{2\pi k_F} V_{nl}^s(E_F, k_y) \qquad (S5)$$

with

$$V_{nl}^s(E_F, k_y) = A\left\{s^2 + (1-s^2)\cos\left(\frac{\theta_L + \theta_0}{\sqrt{1-s^2}}\right)\right\}, \qquad (S6)$$

where $s \equiv k_y/k_F$ indicates the contribution of transverse modes to spin transport, $\theta_L = 2m^*\alpha(v_g)L/\hbar^2$ indicates the angle by which spins precess on a distance $L$, $\theta_0$ indicates any additional arbitrary phase related, e.g., to the final dimension of the ferromagnetic contacts. For precession angles $\theta_L > 2\pi$ the general formula is well approximated by the following equation[7]:

$$V_{nl} = V_{nl}^0 + \frac{A}{\sqrt{2\pi(\theta_L + \theta_0)}} \cos\left(\theta_L + \theta_0 + \frac{\pi}{4}\right), \qquad (S7)$$

which we introduce as eq. (1) in the main text. We use this equation to qualitatively describe the observed signal oscillations, shown in Fig. 4(b) in the main text. Theoretical curves in that figure are however plotted using eq. S5.